\documentclass[fleqn,12pt,twoside]{article}
\usepackage[headings]{espcrc1}

\readRCS
$Id: espcrc1.tex,v 1.2 2004/02/24 11:22:11 spepping Exp $
\usepackage{graphicx}
\usepackage[figuresright]{rotating}

\def\lsim{\raise0.3ex\hbox{$<$\kern-0.75em\raise-1.1ex\hbox{$\sim$}}}
\def\gsim{\raise0.3ex\hbox{$>$\kern-0.75em\raise-1.1ex\hbox{$\sim$}}}

\newcommand{\AmS}{{\protect\the\textfont2
  A\kern-.1667em\lower.5ex\hbox{M}\kern-.125emS}}

\title{Hadronic fluctuations in the QGP} 

\author{F. Karsch\address[BNL]{Physics Department, 
Brookhaven National Laboratory, Upton, NY 11973, USA}\thanks{This work 
was partly supported by the  KBN under grant 2P03
(06925), the DFG under grant KA 1198/6-4 and the GSI collaboration grant BI-KAR.
The work of FK has been partly supported by a contract DE-AC02-98CH1-886 with
the U.S. Department of Energy.
},
S. Ejiri\address{Department of Physics, The University of Tokyo,
Tokyo 113-0033, Japan} 
and
K. Redlich\address{Physics Department, Theory Division, CERN, CH-1211 Geneva 23,
Switzerland}
}

\runtitle{Hadronic fluctuations in the QGP}
\runauthor{F. Karsch, S. Ejiri and K. Redlich}

\begin{document}

\maketitle

\begin{abstract}
We analyze fluctuations of quark number and electric charge, in 2-flavour
QCD at finite temperature and vanishing net baryon number density.
In the hadronic phase we find that an enhancement of charge fluctuations
arises from contributions of doubly charged hadrons to the thermodynamics.
The rapid suppression of fluctuations seen in the high temperature phase
suggests that in the QGP
quark number and electric charge are predominantly carried by quasi-particles
with the quantum numbers of quarks.
\end{abstract}

\section{Introduction}

Lattice calculations of bulk thermodynamic observables, e.g. the 
energy density and pressure, clearly show that the transition to
the high temperature phase of QCD is accompanied by the liberation
of partonic degrees of freedom. Asymptotically, at infinite temperature, energy 
density and pressure approach the limit of an ideal gas of quarks and gluons. 
At $T\sim 3\; T_c$ the deviations from this asymptotic
behaviour are still about 15\% which is too large to be accounted for
by ordinary high temperature perturbation theory. This shows the relevance
of non-perturbative effects leading, for instance, to thermal quark and 
gluon masses. For temperatures close to $T_c$, {\it i.e.} $T\sim (1-2)\; T_c$, 
these non-perturbative features dominate bulk thermodynamic behaviour, 
leading to large deviations from the ideal gas relation $\epsilon = 3p$, a 
strong reduction of the velocity of sound and large screening lengths \cite{review}.
 
The observation of large elliptic flow in heavy ion collisions at RHIC,
its successful description in terms of ideal hydrodynamics as well as the 
observed strong
modification of jets also led to the conclusion that the medium produced in 
heavy ion collisions at RHIC, which is expected to be generated and
thermalized at temperatures $T\sim (1-2)\; T_c$, is opaque and still strongly 
interacting \cite{RHIC}.
It thus became of considerable interest to understand in more detail the
structure of the interacting medium in the vicinity of the transition
temperature. In this context it has been suggested that a large set of
colored bound states of light quarks could exist above $T_c$ and dominate
the bulk thermodynamics for $T \simeq (1-2)\;T_c$ \cite{shuryak}. 

Indeed, lattice calculations of in-medium properties of heavy quark bound states 
have led to the conclusions that correlation functions for some of these 
states, in particular $J/\psi$ and $\eta_c$, are not significantly modified 
in the vicinity of $T_c$ and that these states thus can persist to exist as bound 
states even at $T\sim (1.5-2)\; T_c$ \cite{quarkonium}. 
On the other hand, there is ample 
evidence that correlation functions of light quarks are strongly modified
immediately above $T_c$. For instance, scalar and pseudo-scalar correlation
functions become degenerate as a consequence of chiral symmetry restoration;
in the chiral limit the pseudo-scalar particle no longer is a Goldstone 
particle which is reflected in the structure of spectral functions where the
temperature independent, $\delta$-function like peak present at low temperature 
gets replaced by a broad temperature dependent ''bump''  at energies 
$\omega \sim 5T$ \cite{dileptons}. Moreover, the analysis of spatial correlation 
functions in the scalar and vector channels showed a strong sensitivity to 
modifications of temporal boundary conditions for the 
fermion (quark) fields above $T_c$ while they are insensitive to this below
$T_c$ \cite{screening}. 
This suggests that in these quantum number channels a bosonic bound
state does not give the dominant contribution to the correlation functions
above $T_c$; the fermionic substructure of ''independently'' propagating  
quarks and anti-quarks becomes visible. If light quark bound 
states exist above $T_c$ they thus must have quite peculiar quasi-particle
properties. 

In recent lattice studies of 2-flavour QCD at non-zero quark ($\mu_{u,d}$)  
chemical potential higher order derivatives of the QCD partition function 
(generalized
susceptibilities) with respect to quark chemical potentials have been
used to also analyze higher moments of fluctuations of quark number and 
electric charge \cite{lgt3}. 
In particular, the second order derivatives with respect to $\mu_{q}$ or $\mu_Q$
are monotonously rising across $T_c$ while the fourth order derivatives 
have pronounced peaks at the transition temperature. We will show in the
following that this reflects the transition from hadrons to quarks as the
dominant degrees of freedom that carry baryon number and electric charge
in the hot medium \cite{ours}.  

\section{Quark number and charge fluctuations}  

\begin{figure}[htb]
\begin{center}
\begin{minipage}[t]{55mm}
\includegraphics[width=5.0cm]{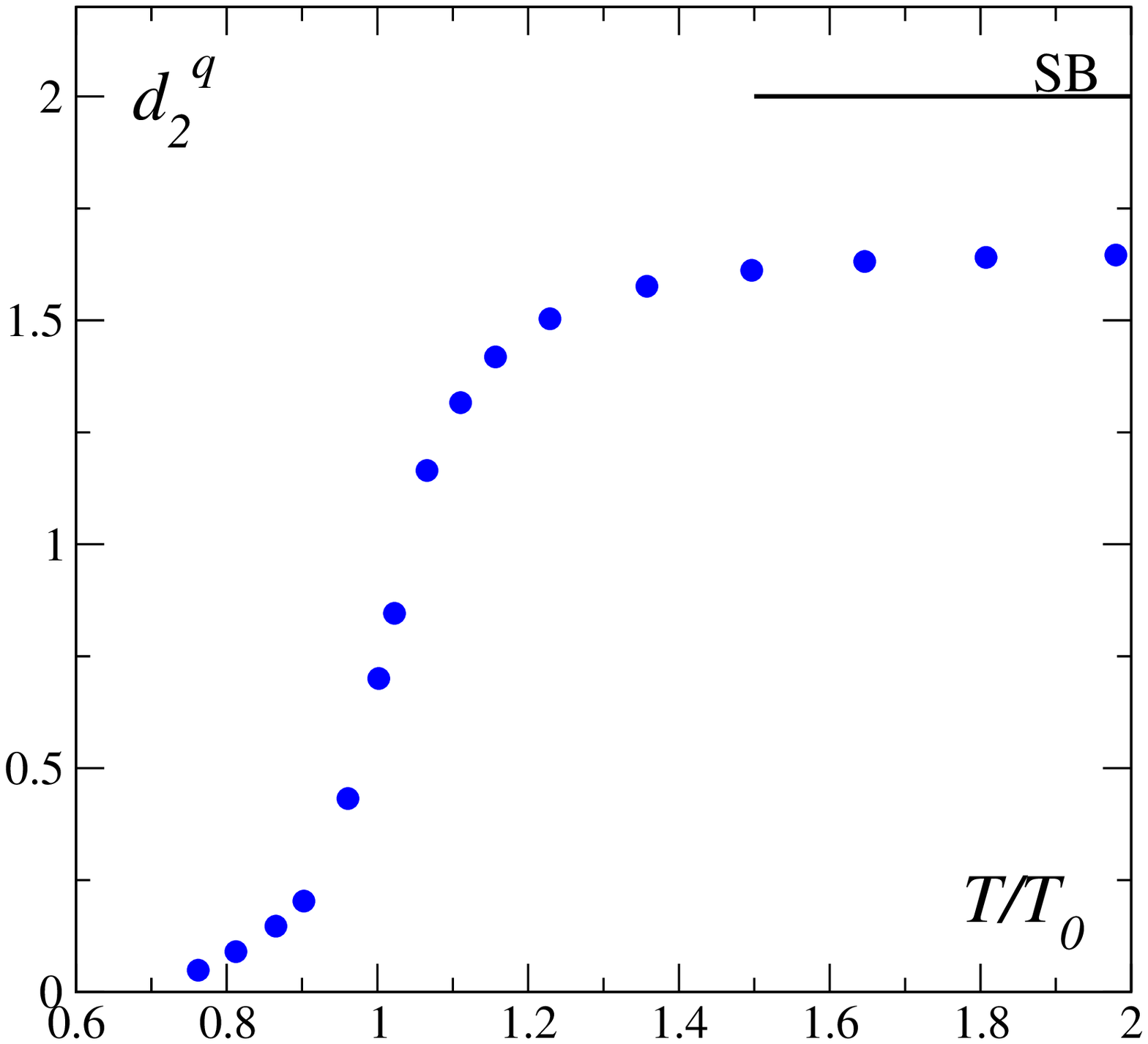}
\end{minipage}
\hskip 0.6cm
\begin{minipage}[t]{55mm}
\includegraphics[width=5.0cm]{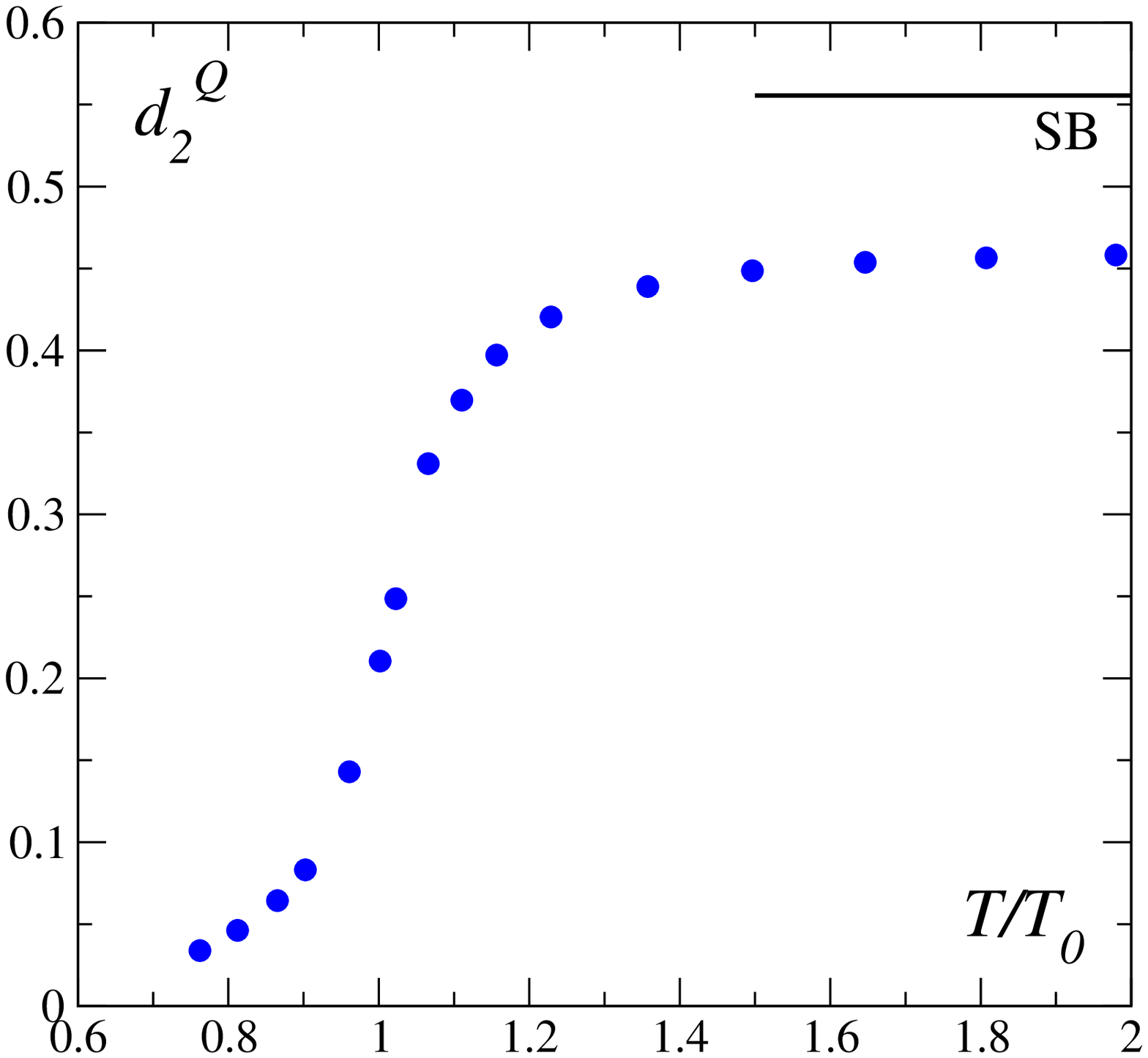}
\end{minipage}
\vskip 0.2cm
\begin{minipage}[t]{55mm}
\includegraphics[width=5.0cm]{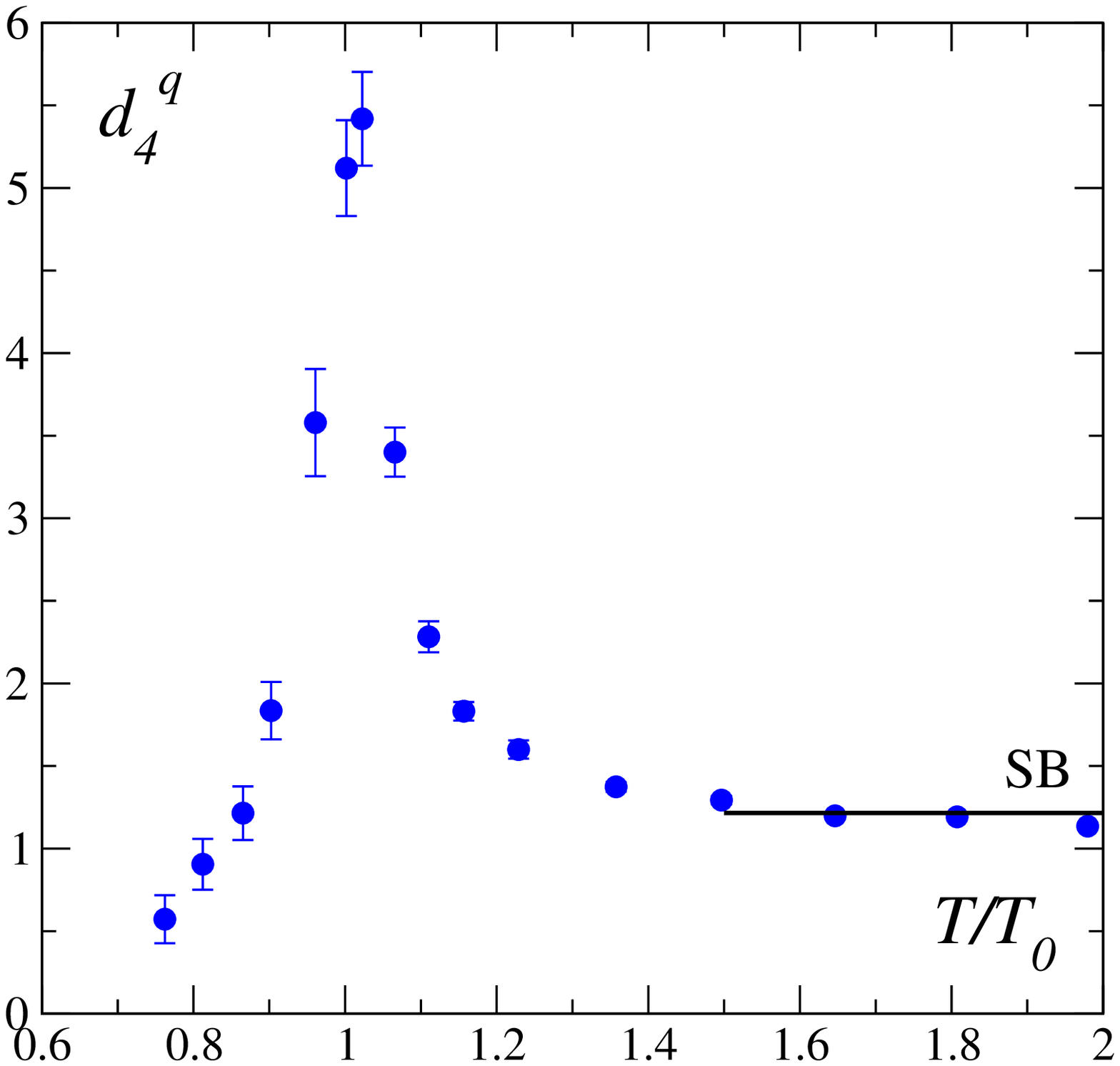}
\end{minipage}
\hskip 0.6cm
\begin{minipage}[t]{55mm}
\includegraphics[width=5.0cm]{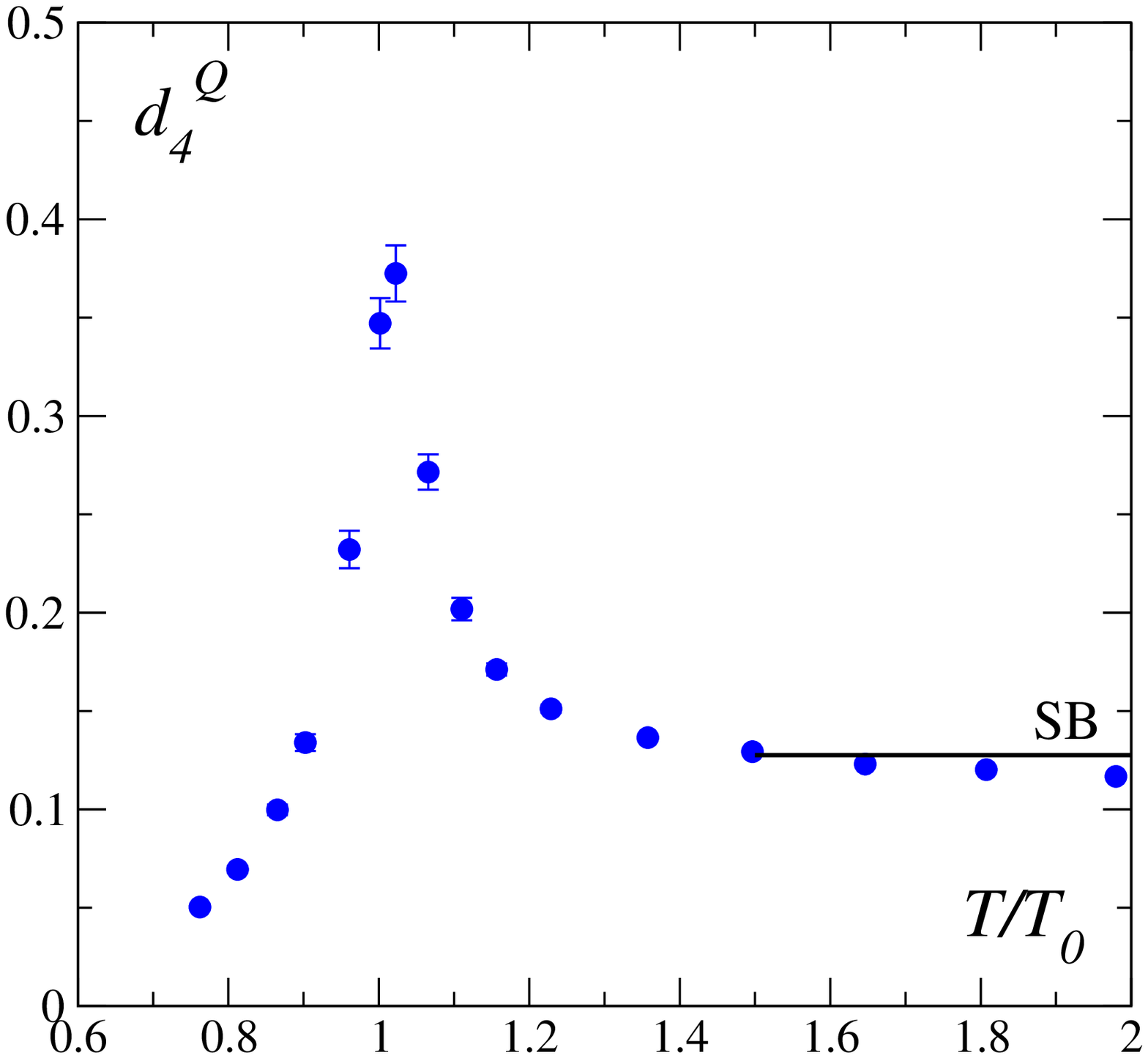}
\end{minipage}
\end{center}

\vskip -1.0cm
\caption{Second (upper) and fourth (lower) cumulant of the net quark number
(left) and electric charge (right) in 2-flavour QCD calculated on lattices
of size $16^3\times 4$ with quark masses corresponding to a pseudo-scalar
(pion) mass of about 770~MeV.} 
\label{fig:data}
\vskip -0.4cm
\end{figure}

Fluctuations of quark number and charge at vanishing net baryon number
density (vanishing chemical potential) can be obtained from the QCD
partition function through appropriate combinations of derivatives with
respect to $u,d$-quark chemical potentials,
\begin{eqnarray}
d_2^x \equiv\frac{1}{VT^3}
\frac{\partial^2 \ln Z}{\partial (\mu_{x}/T)^2}\biggl|_{\mu_u=\mu_d=0} &=& 
\frac{1}{VT^3}
\langle (\delta N_{x})^2 \rangle \quad ,\quad x=q,~Q
\nonumber \\
d_4^x\equiv \frac{1}{VT^3}
\frac{\partial^4 \ln Z}{\partial (\mu_{x}/T)^4}\biggl|_{\mu_u=\mu_d=0} &=& 
\frac{1}{VT^3}
\left( \langle (\delta N_{x})^4 \rangle - 3 \langle (\delta N_{x})^2
\rangle^2 \right)    \quad ,
\label{fluctuations}
\end{eqnarray}
where $\partial/\partial \mu_q= \partial /\partial\mu_{u}+\partial /\partial\mu_{d}$
and $\partial/\partial \mu_Q = \frac{2}{3}\partial/\partial \mu_u -
\frac{1}{3}\partial/\partial \mu_d$. Results for these 
cumulants obtained from a calculation within 2-flavour QCD are shown in 
Fig.~\ref{fig:data}.

The ratios of the quark number and electric charge cumulants, 
$R_{4,2}^x\equiv d_4^x/d_2^x$, $x=q,~Q$,
are sensitive observables that allow to identify the carriers of quark number
and electric charge in a thermal medium. This is quite apparent in the low
temperature phase where a hadron resonance gas (in Boltzmann approximation) is 
known to give a good description of the bulk thermodynamics \cite{Tawfik}. 
This immediately
leads to the expectation that $R_{4,2}^q = 3^2=9$ for $T< T_c$. The analysis
of electric charge fluctuations is a bit more subtle as doubly charged baryons
start contributing to the thermodynamics significantly for $T\simeq T_c$.
Taking into account separately the contribution from the mesonic isospin
triplet sector ($G^{(3)}$) and the baryonic isospin doublet ($F^{(2)}$) 
and quartet ($F^{(4)}$) sectors one finds for the charge fluctuations \cite{ours},
\begin{equation}
R^Q_{4,2} \equiv  \frac{d^Q_4}{d^Q_2} =
\frac{\langle (\delta Q)^4 \rangle}{\langle (\delta Q)^2 \rangle}- 3
\langle (\delta Q)^2 \rangle = {4 G^{(3)} + 3 F^{(2)} + 27 F^{(4)} \over
4 G^{(3)} + 3 F^{(2)} + 9 F^{(4)} }
\quad .
\label{R42Q}
\end{equation}
At low temperature this ratio is dominated by charge fluctuations in
the pion sector, which contributes to $G^{(3)}$. The ratio
will thus approach unity at low temperatures and is monotonically
increasing with temperature. This indicates that in the low temperature
limit all charged degrees of freedom carry one unit of charge;  
$R_{4,2}^Q$ increases only due to contributions arising from isospin 
quartet baryons which can carry charge $Q=2$.

\begin{figure}[htb]
\begin{center}
\includegraphics[width=5cm]{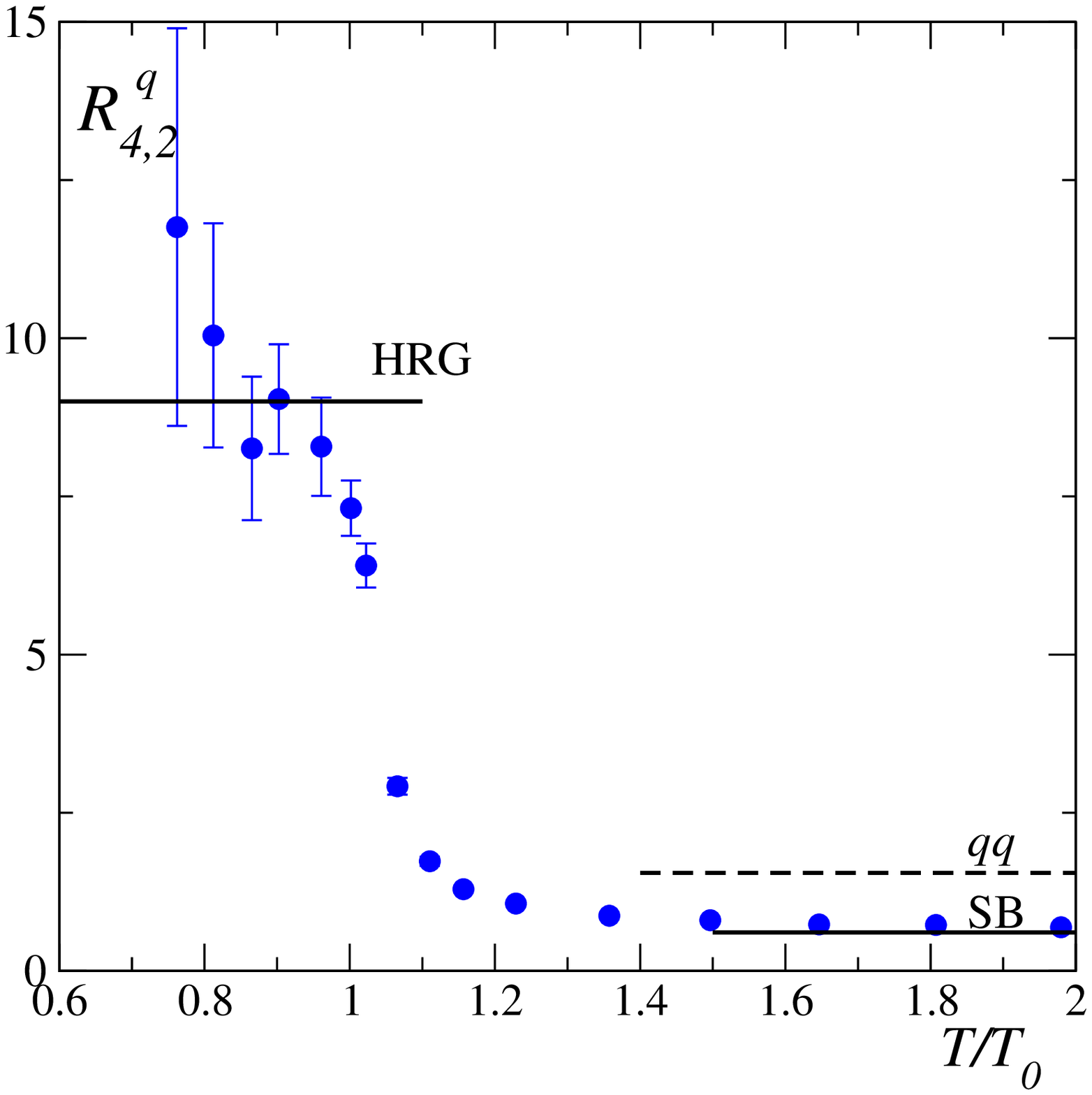}\hspace{0.5cm}
\includegraphics[width=5cm]{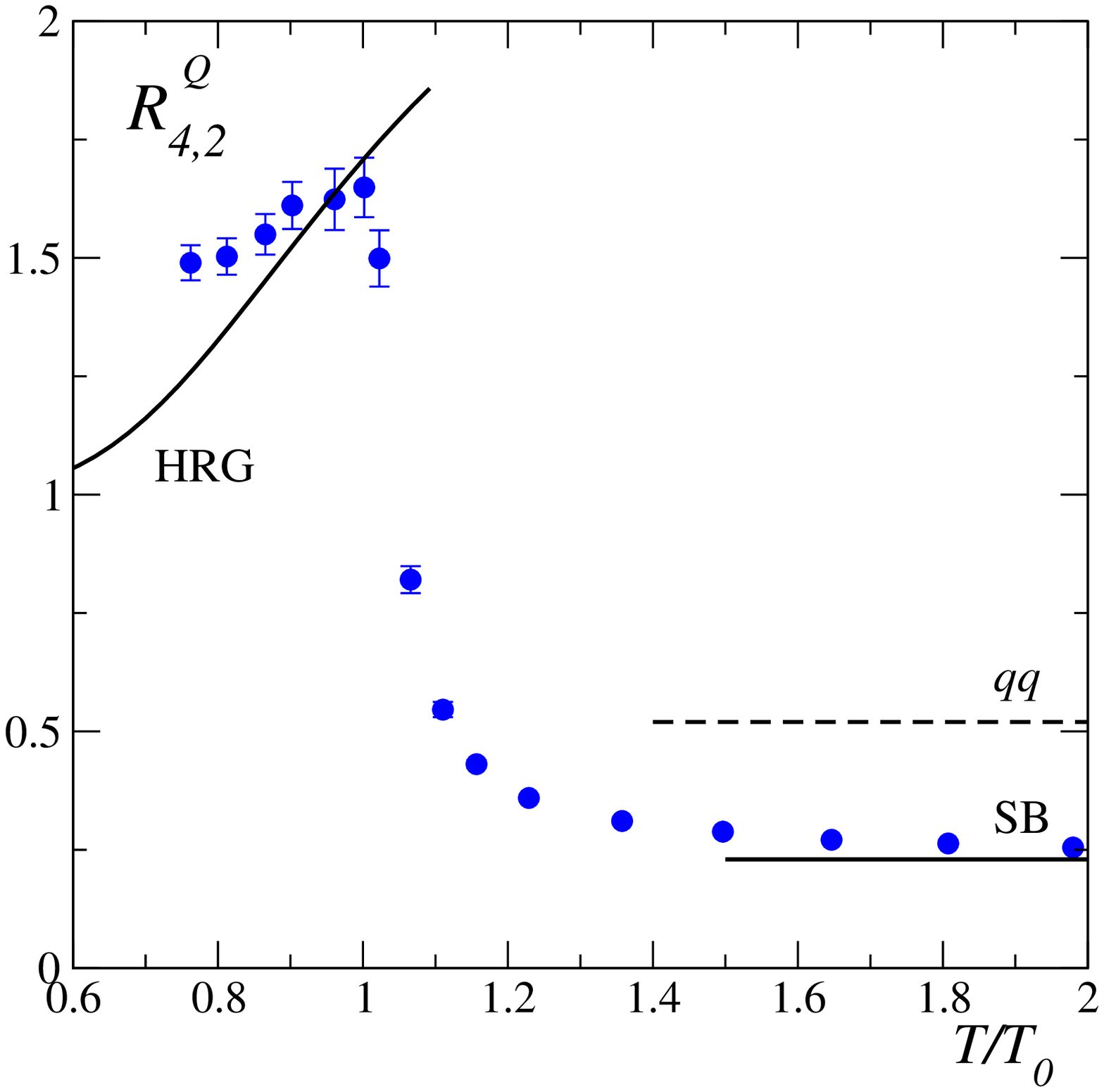}
\end{center}
\vskip -1.0cm
\caption{
The ratios of fourth and second cumulants of quark number (left)
and charge (right) fluctuations.} 
\label{fig:R42}
\vskip -0.4cm
\end{figure}

The generic features expected for quark number and charge fluctuations in a hadron
gas are reproduced by the lattice results for $R_{4,2}^{q,Q}$ 
shown in Fig.~\ref{fig:R42}. In the high temperature phase these ratios
apparently change drastically. In fact, in the infinite temperature, ideal gas 
limit the ratios will approach $R_{4,2}^{q,\infty} = 6/\pi^2$ and 
$R^{Q,\infty}_{4,2} = 34/15\pi^2$, respectively. As can be seen from 
Fig.~\ref{fig:R42} these asymptotic values are almost reached for $T\gsim 1.5\; T_c$.
 
Like in the hadronic phase one would expect also for the high temperature
phase that the presence of doubly charged states or states carrying more 
than one unit of quark number would lead to an increase of the 
ratios $R_{4,2}^{q,Q}$. The analysis of light quark bound states 
\cite{shuryak} indeed suggests that in addition to colored $\bar{q}q$ and $gg$
states also $qq$-states could exist above $T_c$. The latter, however, would 
be only weakly bound and presumably would disappear already at $T\simeq 1.4 T_c$.
If such states would contribute to the thermodynamics above $T_c$ they also
would lead to an increase in charge and quark number fluctuations. The dashed lines
shown in Fig.~\ref{fig:R42} indicate the increase in $R_{4,2}^{q,Q}$ at the presumed 
melting temperature of $qq$-states, $T\simeq 1.4 T_c$ \cite{shuryak}, assuming that at 
this temperature the $qq$-states contribute only with half their statistical weight 
to the QCD partition function. This clearly overestimates the fluctuations observed
at high temperature in lattice calculations.

\section{Conclusions}
                                                                                
We have discussed some generic features of quark number and charge fluctuations 
in 2-flavour QCD. We argue that the ratio of quartic
and quadratic fluctuations are sensitive observables that can directly
provide information on the constituents of the thermal medium that
carry net quark number and electric charge, respectively. We have shown
that below the QCD transition temperature these ratios are in reasonable
agreement with a hadronic resonance gas. Above $T_c$ the ratios
rapidly drop and approach the high temperature ideal gas values. This
suggests that already for $T\gsim 1.5T_c$ quark number and charge are
predominantly carried by states with the quantum numbers of quarks.

\end{document}